\documentclass[%
 reprint,
superscriptaddress,
nobibnotes,
 amsmath,amssymb,
 aps,
 prd,
floatfix,
]{revtex4-2}

\usepackage{booktabs}
\usepackage{graphicx}
\usepackage{dcolumn}
\usepackage{bm}
\usepackage{verbatim}
\usepackage{xcolor}
\usepackage[utf8]{inputenc}
\DeclareUnicodeCharacter{2264}{\ensuremath{\leq}}
\usepackage{hyperref}

\DeclareUnicodeCharacter{200B}{}

\begin{document}

\preprint{APS/123-QED}

\title{Assembly Bias in eRASS1 X-ray Galaxy Clusters: Accounting for Non-uniform Survey Exposure}

\author{Jitendra Joshi}
\affiliation{Inter University Centre for Astronomy and Astrophysics, Ganeshkhind, Pune 411007, India}
\affiliation{Institute of Physics, University of Szczecin, Wielkopolska 15, 70-451 Szczecin, Poland}
\author{Divya Rana}%
\affiliation{Leiden Observatory, Leiden University, PO Box 9513, NL-2300 RA Leiden, The Netherlands}
\author{Surhud More}
\affiliation{Inter University Centre for Astronomy and Astrophysics, Ganeshkhind, Pune 411007, India}
\affiliation{Kavli Institute for the Physics and Mathematics of the Universe (WPI), University of Tokyo, 5-1-5, Kashiwanoha, 2778583, Japan}

\date{\today}

\begin{abstract}
Halo assembly bias describes the dependence of halo clustering on properties beyond halo mass and provides a means of connecting halo assembly history to the large-scale matter distribution. We develop and test a relative clustering methodology for measuring differences in the large-scale bias of two halo populations using the ratio of their cluster--galaxy pair counts. In the linear regime, this ratio can be related to the difference in cluster bias, while the galaxy autocorrelation independently constrains the galaxy bias. We apply the method to a volume-limited X-ray selected galaxy clusters from the eROSITA eRASS1 catalogue, dividing them into high- and low-temperature subsamples with matched distributions in X-ray luminosity, redshift, and exposure. We first verify the validity of the linear approximation, finding $\xi(r)\ll1.0$ over $4$--$10\,h^{-1}\mathrm{Mpc}$. The joint clustering analysis yields $b_{\rm g}=1.610^{+0.018}_{-0.019}$ and a bias difference $\Delta b=0.8\pm1.1$. DES Y3 weak gravitational lensing measurements provide an independent test of the halo masses, yielding $\log(M_{200\mathrm{m}}/h^{-1}M_\odot)=14.33^{+0.07}_{-0.07}$ and $14.33^{+0.10}_{-0.09}$ for the high- and low-temperature samples, respectively, with a mass difference of $\Delta\log M_{200\mathrm{m}}=-0.01\pm0.12$. We therefore find no statistically significant evidence for assembly bias at the $1\sigma$ level, but our measurement is consistent with the bias difference measured for analogous halo populations in the FLAMINGO simulation. The result therefore does not rule out assembly bias at the level expected from the simulation, while demonstrating that the relative clustering methodology provides a framework for future assembly-bias measurements with larger X-ray cluster and galaxy surveys.

\begin{description}
\item[Usage]
Secondary publications and information retrieval purposes.

\end{description}
\end{abstract}

\maketitle

\section{Introduction}
\label{sec:introduction}

Within the framework of the cold dark matter (CDM) paradigm, massive galaxy clusters are expected to form from rare high peaks in the primordial density field. These massive halos act as biased tracers of the underlying matter distribution and therefore exhibit stronger clustering than the matter field itself. The clustering strength of dark matter halos is known to depend primarily on halo mass \citep[e.g.,][]{Kaiser1984, Efstathiou1988, MoWhite1996}. 

Numerical studies, however, have demonstrated that halos of identical mass can display different clustering properties depending on secondary characteristics connected to their formation and assembly histories \citep[e.g.,][]{gao_etal05, wechsler06, gao_white07, faltenbacher_white10, dalal_etal08}. This phenomenon is commonly referred to as \textit{halo assembly bias}. It arises because halos residing in different large-scale environments experience distinct growth histories, which in turn affect both their spatial clustering and internal structure \citep{bullock01,wechsler02,hahn_etal07,faltenbacher_white10}. Consequently, halo clustering can also correlate with structural properties such as halo concentration. Although the magnitude and even the sign of halo assembly bias can depend on the particular definition of the secondary property and on halo mass itself \citep[e.g.,][]{Mao_etal2018,han_etal2019}, many studies have found that, at the massive end, low-concentration halos, typically associated with later formation times, cluster more strongly than high-concentration halos \citep[e.g.,][]{dalal_etal08}.

Given the robust theoretical evidence for halo assembly bias in simulations, considerable effort has been devoted to detecting this effect observationally using galaxies and galaxy groups. However, previous attempts have yielded either weak or inconclusive results \citep[e.g.,][]{yang_etal06a, wang_etal13, Dvornik_etal2017, Niemiec_etal2018, Lin2016}. A major challenge in observational studies is the lack of direct measurements of halo mass and halo formation history, requiring the use of indirect observational proxies. In practice, two broad classes of proxies are commonly employed: those based on the spatial distribution of member galaxies within groups or clusters, and those based on the properties of the central galaxies themselves.

At the scale of galaxy clusters, \citet{Miyatake2016} and \citet{More2016} reported evidence for halo assembly bias using the optically selected redMaPPer cluster catalogue \citep{Rykoff_etal2014}. In their analyses, cluster richness, defined as the membership-probability-weighted number of galaxies, was used as a proxy for halo mass, while the spatial compactness of cluster member galaxies served as a proxy for halo concentration. Weak gravitational lensing measurements showed that the resulting cluster subsamples possessed statistically consistent halo masses, yet exhibited markedly different clustering amplitudes. The observed difference in clustering strength reached nearly $60\%$, substantially exceeding theoretical expectations from $N$-body simulations.

The unexpectedly large signal motivated several follow-up investigations aimed at identifying its origin \citep{Zu2017, BuschWhite2017, SunayamaMore2019}. These studies concluded that the apparent assembly bias signal was most likely driven by optical projection effects. In particular, galaxies projected along the line of sight were incorrectly identified as cluster members, artificially altering the inferred compactness of the galaxy distribution. Since the compactness measure was strongly correlated with the fraction of projected interlopers, the derived concentration proxy became significantly contaminated. As a result, observational proxies constructed from member galaxies in optically selected clusters are intrinsically susceptible to projection-induced systematics.

Such issues related to the projection effects on the measurements on the intermediate to large scales in optically selected galaxy clusters can be avoided by using clusters selected using the Sunyaev Zeldovich (SZ) effect \citep{1970Ap&SS...7....3S, 1980MNRAS.190..413S}. \citet{2019ApJ...874..184Z}  used galaxy clusters selected from the Planck SZ survey and cross-correlated them with galaxies detected in the Pan-STARRS. \citet{2019MNRAS.487.2900S} used SZ clusters selected from the Atacama
Cosmology Telescope (ACT) Polarimeter \citep[][]{2018ApJS..235...20H} and the South Pole Telescope  \citep[SPT,][]{2015ApJS..216...27B} SZ survey along with galaxy catalog data from the Dark Energy Survey \citep[DES,][]{2005astro.ph.10346T}. Furthermore, studies along the same lines have used galaxy clusters identified using the X-rays emitted by the bremsstrahlung emission from ICM \citep{2017ApJ...836..231U, 2019MNRAS.485..408C, 2021ApJ...911..136B} and found similar results.

In this work, we use the galaxy cluster catalogue from the first eROSITA All-Sky Survey (eRASS1) \citep{Bulbul2024}. An important feature of the eRASS1 cluster sample is that the survey exposure is not uniform across the sky. Variations in exposure can affect the sensitivity to X-ray sources and therefore the observed cluster sample. Consequently, when comparing the clustering of two cluster populations, differences in their exposure distributions need to be controlled so that the comparison is not affected by differences in the underlying survey selection. We therefore construct the two populations to have matched distributions in exposure, in addition to matching their X-ray luminosity and redshift distributions. We further use the FLAMINGO cosmological simulations to assess whether the measured relative clustering difference is consistent with the expected assembly-bias signal in simulations.

We develop a relative clustering methodology to search for assembly bias in X-ray selected galaxy clusters. Rather than measuring the absolute clustering amplitude of each cluster subsample independently, we consider the ratio of the cluster--galaxy pair counts for two populations selected from the same parent sample. When the correlation function is sufficiently small, this ratio can be expanded in the linear regime and expressed in terms of the difference in the linear bias of the two cluster populations. The galaxy autocorrelation function provides an independent constraint on the galaxy bias, allowing the relative cluster bias to be inferred from a joint analysis. We apply this approach to the eROSITA eRASS1 cluster catalogue, using X-ray temperature to define two subsamples with matched distributions in luminosity, redshift, and exposure. We then use DES Year 3 (Y3) weak gravitational lensing measurements to verify that the two subsamples have statistically consistent halo masses, providing an essential consistency check for interpreting a relative clustering difference as a possible signature of assembly bias.

The relative nature of our approach also reduces the dependence on model-based assumptions involved in the construction of random catalogues for the eRASS1 cluster sample. In the clustering analysis of \citet{Seppi2024}, the cluster random catalogue is constructed using the eRASS1 selection function, with cluster-property distributions derived from the digital-twin simulations \citep{Seppi2022,Seppi2024}. The selection function accounts for variations in count rate, redshift, exposure, background, and Galactic absorption, while the underlying digital-twin simulation adopts a model for the cluster population. Such a construction provides a detailed treatment of the survey selection, while incorporating assumptions from the underlying cluster and detection models. This is relevant for clustering measurements, as \citet{Seppi2024} note that a traditional sensitivity-map-based construction \citep{Georgakakis2008} may introduce spurious clustering for extended X-ray sources. However, the random catalogue generated in \citet{Seppi2024} is not publicly available, which limits the direct reproducibility of clustering analyses requiring such a catalogue using publicly available eROSITA cluster data. In contrast, our method uses two populations drawn directly from the observed eRASS1 catalogue, with their luminosity, redshift, and exposure distributions explicitly matched, providing a more directly data-driven comparison of their relative clustering without requiring the construction of an independently modelled cluster random catalogue.

The different data catalogs employed in this study are described in Sec.~\ref{sec:data}, whereas the measurement procedures and modeling approaches are presented in Sec.~\ref{sec:modeling}. The primary results of our assembly bias analysis are discussed in Sec.~\ref{sec:results}, followed by a summary of the main findings in Sec.~\ref{sec:conclusion}. Throughout the analysis, we adopt a flat $\Lambda$CDM cosmological model with matter density parameter $\Omega_{\mathrm{m}} = 0.27$, baryon density $\Omega_{\mathrm{b}} = 0.049$, scalar spectral index $n_{\mathrm{s}} = 0.95$, fluctuation amplitude $\sigma_{8} = 0.81$, and dimensionless Hubble parameter $h = 0.7$. Here, $r$ denotes the three-dimensional radial distance from the cluster center, while $R$ represents the corresponding projected two-dimensional radial distance. We define the halo mass using $M_{200\mathrm{m}}$, with the associated halo boundary $R_{200\mathrm{m}}$ corresponding to the radius within which the mean matter density is 200 times the present-day matter density of the Universe. Throughout the paper, $\log$ denotes the base-10 logarithm.

\section{Data}
\label{sec:data}
In this section, we describe the data used for weak gravitational lensing and cluster-galaxy cross correlation analysis.
\subsection{Galaxy Cluster Catalog}
The extended ROentgen Survey with an Imaging Telescope Array (eROSITA; \cite{predehl_erosita_2021}) is the primary instrument onboard the Spectrum–Roentgen–Gamma (SRG) mission \cite{Sunyaev_SRG_2021}, operating at the Sun–Earth L2 Lagrange point. eROSITA consists of seven identical Wolter-I mirror modules and is sensitive in the $0.2$–$8,\mathrm{keV}$ X-ray energy range. During its nominal mission, eROSITA has completed four all-sky surveys \cite{Countinho_2022}.

In this work, we use the galaxy cluster catalog constructed in the soft X-ray band ($0.2$–$2.3,\mathrm{keV}$) from the first eROSITA All-Sky Survey (eRASS1) in the western sky \citep{Bulbul2024}. This catalog contains 5,259 optically confirmed galaxy clusters over a sky area of $12{,}791,\mathrm{deg}^2$, with cluster properties derived from X-ray measurements and optical counterpart identification.

We construct an approximately volume-limited sample by applying cuts in X-ray luminosity and redshift, selecting clusters with $L_{500} > 10^{43.5}\,\mathrm{erg\,s^{-1}}$ and $0.15 < z_{\rm red} < 0.65$. To ensure a common survey footprint for clustering measurements, we further restrict the sample to the sky region overlapping with the DES Y3 galaxy catalogs. We additionally require clusters to have reliable intracluster medium temperature measurements derived using the MBProj2D method \citep{Sanders2018}. After applying these selection criteria, the final sample consists of 524 galaxy clusters, which we use for our assembly bias analysis.

The resulting cluster sample is well suited for large-scale structure analyses, as the X-ray selection provides a clean and physically motivated tracer of massive dark matter halos. The wide sky coverage and redshift range of the eRASS1 western-sky sample along with overlap with DES survey enable robust measurements of cluster–galaxy correlations on large scales.

\subsection{\label{sec:galaxy catalog}Galaxy Catalogs}

We use galaxy catalogs from the DES \citep{darkenergysurvey}, a $\sim5000\,\mathrm{deg}^2$ optical imaging survey of the southern sky conducted with the DECam instrument in Chile. DES provides photometric observations in the $g$, $r$, $i$, $z$, and $Y$ bands, reaching typical magnitude limits of $g \simeq 24.3$, $r \simeq 24.1$, $i \simeq 23.6$, $z \simeq 22.8$, and $Y \simeq 21.7$. 

For this work, we use two complementary galaxy samples derived from DES Y3 data: the red-sequence Matched-filter Galaxy Catalog (RedMaGiC) constructed from the DES Gold catalog, and the DES Y3 shape catalog.

\subsubsection{\label{sec:RedMaGiC}DES Y3 RedMaGiC galaxy catalog}

The assembly bias is measured using galaxies selected with the RedMaGiC algorithm \cite{Rozo2016} applied to DES Y3 data. RedMaGiC fits each galaxy to red-sequence templates to estimate a photometric redshift $z_{\rm photo}$ and corresponding luminosity $L$. The catalog \cite{Pandey2021} is constructed by selecting bright galaxies ($L \ge L_{\rm min}$), where $L_{\rm min}=0.5,L_*$ for the high-density sample, with good red-sequence fits ($\chi^2/{\rm dof} \le 2$). We use the high-density RedMaGiC sample, which is volume-limited to a $z$-band luminosity of $0.5\,L_*$, has a limiting magnitude of $m_z = 21.3$, and is divided into three tomographic redshift bins of $z \in (0.15,0.3]$, $(0.3,0.5]$, and $(0.5,0.65]$.

\subsubsection{\label{sec:desy3} DES Y3 Shape Catalog}

The DES Y3 shape catalog contains over 100 million galaxies with measured shapes derived using the \textsc{metacalibration} technique described in \citet{Huff2017} and \citet{Sheldon2017}. 

In DES Y3, \textsc{metacalibration} estimates galaxy shapes using imaging in the $riz$ bands. The resulting shear catalog has an effective source number density of $5.59\,\mathrm{arcmin}^{-2}$ and a shape noise of $0.261$. Each galaxy shape is characterized by a two-component ellipticity, $e$, as defined in \citet{Gatti2021}. The ensemble average of these ellipticities provides an estimate of the gravitational shear, $\gamma$, through the relation
\begin{equation}
\langle \gamma \rangle = \langle \mathcal{R} \rangle^{-1} \langle e \rangle\,,
\end{equation}
where $\mathcal{R}$ denotes the shear response matrix and the angled brackets represent averages over the galaxy sample.

The \textsc{metacalibration} framework also accounts for selection effects via an additional response term, $\mathcal{R}_{\mathrm{s}}$, which captures the dependence of measured shapes on the sample selection. The total response is therefore given by $\mathcal{R} + \mathcal{R}_{\mathrm{s}}$, formally a $2 \times 2$ matrix, but in practice well approximated by the mean of its diagonal components (e.g., \citealt{Prat2018}).

For the purposes of this work, which focuses on assembly bias through large-scale clustering measurements, we neglect percent-level multiplicative shear biases (e.g., due to blending; \citealt{Sheldon2020, MacCrann2021}), as they are subdominant compared to the statistical uncertainties in our analysis.


\section{Modelling and Measurement}
\label{sec:modeling}

In this section, we measure halo assembly bias in galaxy clusters by comparing the large-scale clustering of cluster subsamples that are matched in mass-related observables but differ in their X-ray temperature. We divide the cluster catalogue into two subsamples, low- and high-temperature galaxy clusters. The clustering properties of each subsample are measured relative to a common reference galaxy population of DES RedMaGiC galaxies.

\subsection{Galaxy clustering profile}

We quantify the large-scale clustering of galaxy clusters using the cluster--galaxy cross-correlation function, measured with the Davis--Peebles estimator. Let $D_i$ denote the data catalogue of galaxy clusters in subsample $i$, $D_{\rm g}$ the data catalogue of RedMaGiC galaxies, and $R$ a random catalogue constructed to match the angular mask and selection function of the cluster sample.

The Davis--Peebles estimator for the cross-correlation function is given as
\begin{equation}
\xi_i(R) = \frac{D_i D_{\rm g}}{D_i R} - 1 ,
\end{equation}
where $D_i D_{\rm g}(r)$ is the number of cluster--galaxy pairs between subsample $i$ and the galaxy catalogue, and $D_i R$ is the number of cluster--random pairs, both counted in bins of comoving separation $R$. Here, the pair counts are normalized by the corresponding subsample catalogue sizes.


Since both cluster subsamples are drawn from the same parent catalogue and are constructed to have matched redshift and observational selection functions, they share the same random catalogue. The common random normalization therefore cancels when taking the ratio, leaving the ratio of the normalized cluster--galaxy pair counts,
\begin{equation}
\frac{1 + \xi_1}{1 + \xi_2} = \frac{D_1 D_{\rm g}}{D_2 D_{\rm g}} .
\end{equation}

On sufficiently large scales, where clustering is in the linear regime and $|\xi| \ll 1$, this ratio can be expanded to first order as 
\begin{equation}
(1+\xi_1-\xi_2) \approx \frac{D_1D_{\rm g}}{D_2D_{\rm g}} .
\end{equation}

The validity of this approximation for the scales used in our analysis is tested in Appendix~\ref{app:linear_regime}. In the linear bias theory, the cluster-galaxy cross-correlation function can be written as
\begin{equation}
\xi_{\rm cg}(r) = b_{\rm c} \, b_{\rm g} \, \xi_{\rm mm}(r),
\end{equation}
where $b_{\rm c}$ and $b_{\rm g}$ are the linear bias parameters of the cluster and galaxy samples, respectively, and $\xi_{\rm mm}(R)$ is the matter autocorrelation function. On large scales, the relative clustering between the two cluster subsamples can be written as
\begin{equation}
1 + b_{\rm g} \, \Delta b \, \xi_{\rm mm}(R)
\simeq
\frac{D_1 D_{\rm g}}{D_2 D_{\rm g}} ,
\label{eq:bias}
\end{equation}
where $\Delta b = b_1 - b_2$ quantifies the difference in linear bias between the two cluster subsamples, and $b_{\rm g}$ is the linear bias of the galaxy sample.A statistically significant non-zero value of $\Delta b$ would therefore provide evidence for differential halo bias between the two subsamples, which we interpret as an observational signature of assembly bias when their halo masses are consistent.

The galaxy autocorrelation function is modelled as
\begin{equation}
\xi_{\rm gg}(R) = b_{\rm g}^2 \, \xi_{\rm mm}(R) ,
\label{eq:gg_bias}
\end{equation}
which allows us to constrain the galaxy bias $b_{\rm g}$ independently.

We perform a joint modeling of Eqs.~(\ref{eq:bias}) and (\ref{eq:gg_bias}), with $b_{\rm g}$ and $\Delta b$ as free parameters. The joint analysis is required because the relative cluster--galaxy clustering signal constrains the combination $b_g \Delta b$, while the galaxy autocorrelation provides an independent constraint on $b_g$. The projected matter correlation function $\xi_(R)$ is computed by integrating the three-dimensional matter correlation function $\xi_(r)$ along the line-of-sight coordinate $\pi$ for a fixed fiducial cosmology, and is given by
\begin{equation}
    \xi_{\rm mm}(R) = \frac{2}{\pi_{\rm max}} \int_0^{\pi_{\rm max}} \xi_{\rm mm} ( r,z) \, d \pi,
\end{equation}
where the $r = \sqrt{R^2+\pi^2}$ and $\xi_{\rm mm}(r)$ is evaluated at the median redshift $z$ of our cluster sample. We use the public python package \textsc{AUM} \citep{2013_bosch,2013_More, 2013_Cacciato, aum} for the model prediction of the $\xi_{\rm mm}(R)$. We fix $\pi_{\rm max}$ to $200 \, h^{-1} {\rm Mpc}$ to reduce the impact of residual redshift-space and finite line-of-sight projection effects. This choice is motivated by previous studies showing that values of $\pi_{\rm max}\sim150$--$200 \, h^{-1} {\rm Mpc}$ are sufficient to capture the projected clustering signal, with smaller integration limits potentially leading to a systematic loss of signal \citep{Phleps2006,Allevato2016}. The derivation of the relations between the projected pair counts and the linear bias parameters, including the galaxy autocorrelation, is given explicitly in Appendix~\ref{app:mathematical_model}.

\subsection{Galaxy-galaxy lensing profile}
Gravitational lensing arises from the deflection of light emitted by distant background galaxies due to the gravitational potential of intervening matter along the line of sight. In the weak lensing regime, these distortions are small and coherent, and can only be detected statistically through ensemble measurements of galaxy shapes (see \cite{2015_Kilbinger, Mandelbaum2018, dodelson2017book} for reviews). The primary observable is the shear, a complex quantity describing the anisotropic distortion of galaxy images. The tangential shear component, defined relative to the lens–source separation vector, is given by
\begin{equation}
\gamma_t = -\gamma_1 \cos 2\phi - \gamma_2 \sin 2\phi,
\end{equation}
where $\phi$ is the position angle of the source galaxy with respect to the $x$-axis, and $\gamma_1$ and $\gamma_2$ denote the Cartesian components of the shear.

The azimuthally averaged tangential shear is directly related to the projected surface mass density via
\begin{equation}
\langle \gamma_t \rangle(R) = \frac{\overline{\Sigma}(R) - \langle \Sigma(R) \rangle}{\Sigma_{\rm crit}} \equiv \frac{\Delta\Sigma(R)}{\Sigma_{\rm crit}},
\end{equation}
where $\Delta\Sigma(R)$ is the excess surface mass density (ESD). The mean surface density within a projected radius $R$ is defined as
\begin{equation}
\overline{\Sigma}(R) = \frac{\int_0^R \Sigma(R')\, 2\pi R' \, dR'}{\pi R^2},
\end{equation}
while $\langle \Sigma(R) \rangle$ denotes the azimuthally averaged surface density at radius $R$. The critical surface mass density is given by
\begin{equation}
\Sigma_{\text{crit}} = \frac{c^2}{4\pi G} \frac{D_a(z_s)}{(1+z_l)^2 D_a(z_l)\, D_a(z_l, z_s)},
\end{equation}
where $D_a(z_s)$, $D_a(z_l)$, and $D_a(z_l, z_s)$ are the angular diameter distances to the source, to the lens, and between the lens and source, respectively. The factor $(1+z_l)^2$ accounts for the use of comoving coordinates.

We estimate $\Delta\Sigma(R)$ for our lens sample using the estimator
\begin{equation}
\Delta\Sigma(R) = \frac{\sum_{ij} s^{ij} e_t^{ij}(R)}{\sum_{ij} s^{ij} \Sigma_{\mathrm{c,MC}}^{-1}(z_l^i, z_s^j)\, (\mathcal{R}^j + \mathcal{R}_s)},
\end{equation}
where indices $i$ and $j$ label lenses and sources, respectively. Here, $\mathcal{R}$ is the shear response derived from \textsc{metacalibration}, $\mathcal{R}_s$ denotes the selection response, and the weights are defined as
\begin{equation}
s^{ij} = \omega^j \, \Sigma_{\mathrm{c,mean}}^{-1}(z_l^i, z_s^j),
\end{equation}
with $\omega^j$ representing the inverse variance of the shear measurement for each source \cite{Gatti2021}. The inverse critical surface density $\Sigma_{\mathrm{c,MC}}^{-1}$ is computed using Monte Carlo realizations of the source redshift drawn from the BPZ posterior, while $\Sigma_{\mathrm{c,mean}}^{-1}$ is evaluated at the mean BPZ redshift. The impact of photometric redshift uncertainties is at the percent level \cite{McClintock2018}, and is subdominant to current statistical uncertainties.

The stacked weak lensing signal is measured in {\bf 6} logarithmically spaced radial bins over the range $R \in [{\bf 0.1, 1.0}]\, h^{-1}\mathrm{Mpc}$. The cross-component of the signal,
\begin{equation}
\Delta\Sigma_\times = \gamma_\times \Sigma_c,
\end{equation}
is expected to vanish by parity and is therefore used as a null test for residual systematics \cite{Schneider2005}.

The halo boundary is defined by the radius $r_{200\mathrm{m}}$, corresponding to a mean enclosed density of 200 times the present-day matter density of the Universe:
\begin{equation}
r_{200\mathrm{m}} = \left( \frac{3 M_{200\mathrm{m}}}{4\pi \, 200 \rho_m} \right)^{1/3}.
\end{equation}

To model the weak lensing signal, we adopt the Navarro--Frenk--White (NFW) density profile \cite{NFW_profile},
\begin{equation}
\rho_{\mathrm{nfw}}(r) = \frac{\delta_c \rho_m}{(r/r_s)\,(1 + r/r_s)^2},
\end{equation}
with
\begin{equation}
\delta_c = \frac{200}{3} \frac{c^3}{\ln(1+c) - c/(1+c)}, \qquad r_s = \frac{r_{200\mathrm{m}}}{c}.
\end{equation}

This model is fully specified by the parameter set $\Theta = (M_{200\mathrm{m}}, c)$, corresponding to the halo mass and concentration. The predicted $\Delta\Sigma(R)$ is computed using the analytic expressions of \citet{wright1999}, and flat priors are adopted for parameter inference.

We neglect additional contributions such as baryonic point-mass terms, halo miscentering, and explicit 1-halo/2-halo decompositions \cite{Kobayashi2015, Hikage2013, Miyatake2016, Johnston2007}, as these effects are not expected to significantly affect our results on the scales considered. Similarly, we do not employ halo occupation distribution modeling \cite{Seljak2000, Cooray2002, 2013_bosch}, and instead interpret our measurements in terms of an effective halo mass.

\subsection{\label{sec:covariance}Covariance estimation}
We estimate the covariance of our measurements using the jackknife resampling technique \citep{Miller_1974}, which accounts for spatial variations in the data and provides a computationally efficient estimate of uncertainties. The survey footprint is divided into $N=20$ spatial regions, and measurements are recomputed by excluding one region at a time.

The covariance matrix is then given by
\begin{equation}
    C_{ij} = \frac{N - 1}{N} \sum_{k=1}^{N} \left(\mathcal{X}_k^i - \langle \mathcal{X}^i \rangle \right)\left(\mathcal{X}_k^j - \langle \mathcal{X}^j \rangle \right),
\end{equation}
where $i,j$ denote radial bins, $\mathcal{X}_k$ is the measurement from the $k$th jackknife realization, and $\langle \mathcal{X} \rangle$ is the mean over all realizations. In this work, $\mathcal{X}$ corresponds to the cluster--galaxy cross-correlation function.

 We use eqn. 17 in \citet{Hartlap2007}, in order to account for the bias in the inverse covariance matrix due to a finite number of jackknife samples. We obtain the correlation matrix by normalizing each element of our covariance matrix $C_{ij}$ with the diagonal uncertainties
\begin{equation}
    r_{ij} = \frac{C_{ij}}{\sqrt{C_{ii} C_{jj}}}.
\end{equation}
where subscripts ij represent the $\text{i}^{\rm th}$ and $\text{j}^{\rm th}$ radial bins.

At the large projected separations considered in this work, the cluster--galaxy cross-correlation signal has relatively low signal-to-noise, leading to increased bin-to-bin statistical fluctuations arising from finite pair counts and jackknife resampling. Following the procedure adopted by \cite{Mandelbaum_etal2013}, we apply a mild boxcar smoothing filter with length of two bins in radius to suppress these high-frequency fluctuations prior to model fitting. The smoothing is intended only as a numerical regularization of the measured profile and is not expected to modify its underlying large-scale behaviour.

\subsection{\label{sec:mcmc} Model Fitting}

Given the data vector $\mathcal{D}$, we perform a Bayesian inference to estimate the posterior distribution of the model parameters $\Theta = \{b_{\rm g}, \Delta b\}$. We adopt flat priors on both parameters over physically motivated ranges.

Using Bayes' theorem, the posterior distribution can be written as
\begin{equation}
P(\Theta \mid \mathcal{D}) \propto P(\mathcal{D} \mid \Theta)\, P(\Theta),
\end{equation}
where $P(\mathcal{D} \mid \Theta)$ denotes the likelihood of the data given the model parameters, and $P(\Theta)$ represents the prior.

We assume a Gaussian likelihood of the form
\begin{equation}
P(\mathcal{D} \mid \Theta) \propto \exp\left(-\frac{\chi^2(\Theta)}{2}\right),
\end{equation}
with
\begin{equation}
\chi^2(\Theta) =
\left[ \mathcal{D} - \mathcal{M}(\Theta) \right]^T
C^{-1}
\left[ \mathcal{D} - \mathcal{M}(\Theta) \right],
\end{equation}
where $\mathcal{M}(\Theta)$ is the model prediction vector corresponding to the parameters $\Theta$, and $C$ is the covariance matrix of the measurements.

In our analysis, the data vector $\mathcal{D}$ consists of the measured galaxy autocorrelation function, $\xi_{\rm gg}(r)$, and the relative clustering signal, $1 + (\xi_1 - \xi_2)$. The corresponding model vector $\mathcal{M}(\Theta)$ is constructed using Eqs.~(\ref{eq:gg_bias}) and (\ref{eq:bias}), with the matter correlation function $\xi_{\rm mm}(r)$ computed for fiducial cosmology.

We perform a joint fit to both observables in order to simultaneously constrain $b_{\rm g}$ and $\Delta b$, thereby breaking the degeneracy present when considering the relative clustering signal alone.

To sample the posterior distribution, we employ the affine-invariant Markov Chain Monte Carlo (MCMC) ensemble sampler of \citet{Goodman2010}, as implemented in the \textsc{emcee} package \citep{emcee}. We initialize multiple walkers in the parameter space and evolve the chains until convergence, which is assessed through standard diagnostics such as chain stability and autocorrelation times.

The resulting posterior samples are used to infer the constraints on the galaxy bias $b_{\rm g}$ and the assembly bias parameter $\Delta b$.

\section{Results}
\label{sec:results}
This section presents the result of assembly bias measurement in galaxy clusters. We investigate assembly bias using X-ray temperature as a proxy for halo assembly history. At fixed mass, differences in temperature may reflect variations in recent accretion history or dynamical state, which can in turn affect the cluster--galaxy pair counts. It is however essential to control for temperature correlated observables such as luminosity, redshift, and exposure time. The methodology used to achieve this is mentioned in the first part of this section, followed by the weak lensing analysis to show that the subsamples have identical masses. The last part of this section shows the measured assembly bias signal from cross-correlation analysis. 

\subsection{Sample splitting}

\begin{figure*}
    \centering
    \includegraphics[width=\linewidth]{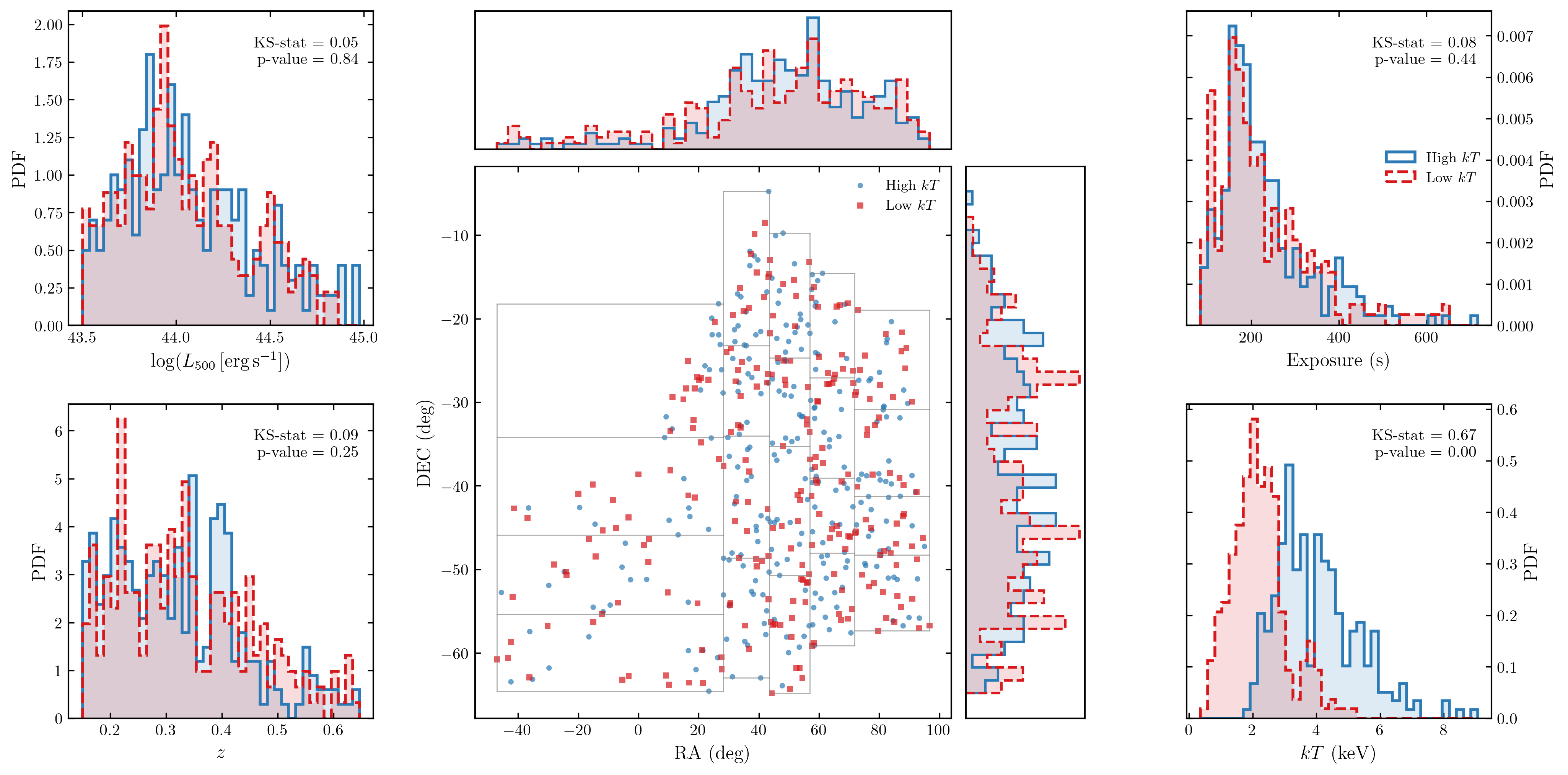}
    \caption{Distributions of the high-$kT$ (blue, solid, circles) and low-$kT$ (red, dashed, squares) cluster subsamples used in the assembly-bias analysis. The upper-left and lower-left panels show the distributions of X-ray luminosity, $\log(L_{500})$, and redshift, $z_{\mathrm{red}}$, respectively, while the upper-right and lower-right panels show the exposure time and temperature, $kT$. The central panel shows the sky distribution of the two subsamples in right ascension (RA) and declination (DEC), with the corresponding marginal RA and DEC distributions shown above and to the right. The KS statistics and p-values indicate that the two subsamples have statistically consistent distributions in luminosity, redshift, and exposure, while the significant difference in $kT$ reflects the temperature-based split.}
    \label{fig:cluster_lum_zred}
\end{figure*}

We adopt a quantile-based splitting scheme to construct low- and high-temperature subsamples while matching their distributions in X-ray luminosity, redshift, and exposure. The full sample is first divided into five logarithmic quantile bins in $L_{500}$. Within each luminosity bin, clusters are further split into three quantile bins in redshift, and each of these is divided into two quantile bins in exposure time. This procedure defines a set of bins in $(\log L_{500}, z_{\mathrm{red}}, E)$ space.

The inclusion of exposure in this matching is important because the eRASS1 survey has a non-uniform exposure across the sky. Since the X-ray sensitivity of the survey depends on the local exposure, the probability of detecting clusters and the precision with which their X-ray properties are measured can vary across the footprint. Matching the exposure distributions ensures that the high- and low-temperature samples are subject to comparable X-ray survey selection, preventing exposure-dependent selection from being coupled to the temperature split.

Within each bin, clusters are separated into low- and high-temperature subsamples using the median temperature of that bin. By construction, this ensures that the resulting subsamples have matched distributions in luminosity, redshift, and exposure, and contain equal numbers of clusters above and below the median temperature. In figure \ref{fig:cluster_lum_zred}, we can see the KS-statistics and the p-values of our subsamples, where the luminosity, redshift and exposures are similar. The sky distribution of the subsamples is also similar. The resulting split has 213 and 211 clusters in high and low temperature subsamples.

\subsection{Weak lensing constraints on halo mass}

\begin{figure*}
    \centering
    \includegraphics[width=0.9\linewidth,keepaspectratio]{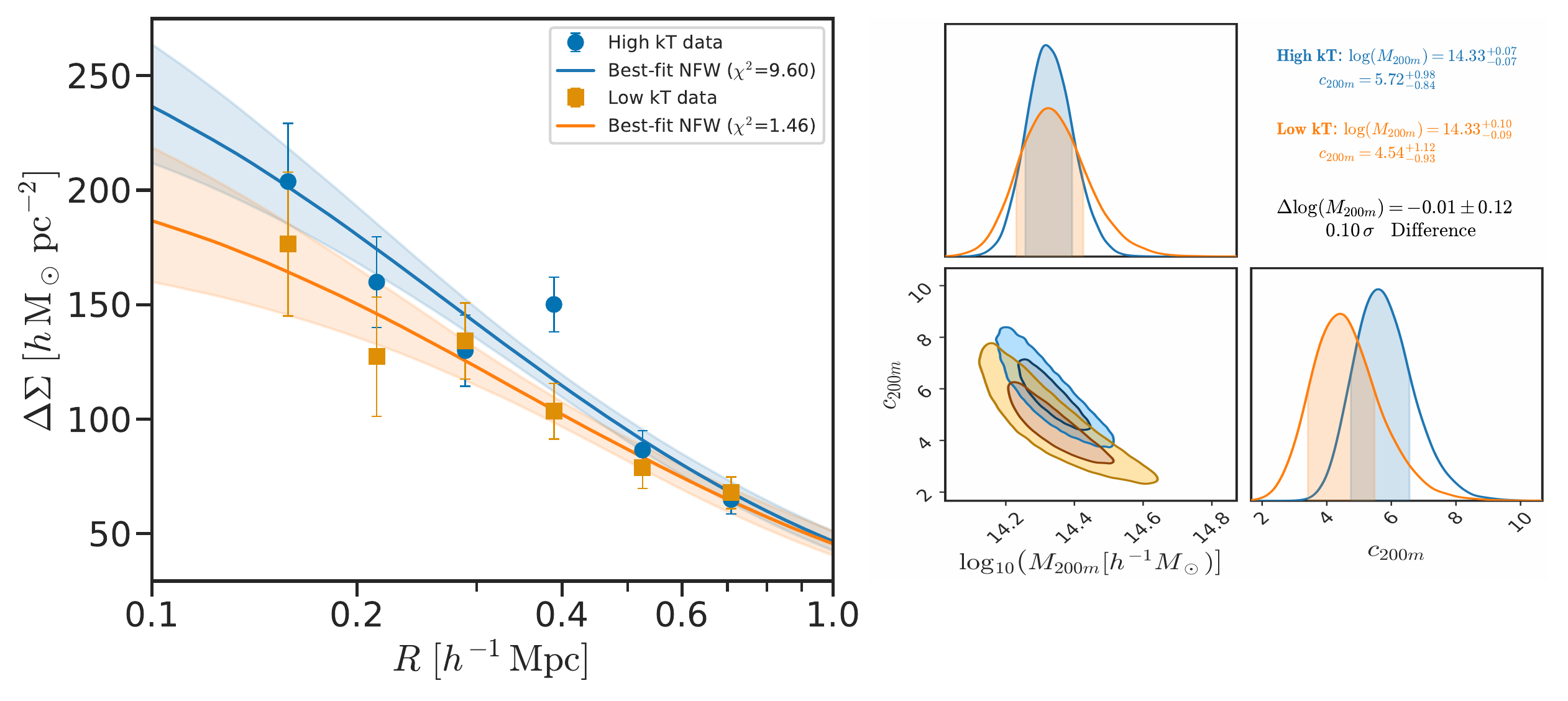}
    \caption{\textit{Weak lensing signal and posterior}: Weak lensing constraints on the halo masses of the high- and low-temperature cluster subsamples. Left: Stacked excess surface density profiles, $\Delta\Sigma(R)$, with the best-fit NFW models and their $1\sigma$ credible regions. Right: Posterior distributions of the fitted halo mass, $\log(M_{200\mathrm{m}}/h^{-1}M_\odot)$, and concentration, $c_{200\mathrm{m}}$. Both subsamples yield statistically consistent halo masses, with $\Delta\log M_{200\mathrm{m}}=-0.01\pm0.12$, indicating that the X-ray temperature split does not introduce a measurable halo mass bias.  }
    \label{fig:WL_result}
\end{figure*}

We measure the stacked weak gravitational lensing signal around the high- and low-temperature cluster subsamples using the DES Y3 shape catalog. The excess surface density profile, $\Delta\Sigma(R)$, is measured in logarithmically spaced projected radial bins over the range $0.1 \leq R \leq 1.0\,{\rm Mpc}$, following the methodology described in Section~\ref{sec:desy3}. The resulting measurements for the two subsamples are shown in Fig.~\ref{fig:WL_result}, together with the best-fit NFW model predictions.

The weak lensing profiles of the two cluster samples exhibit very similar amplitudes over the full radial range considered, indicating that the two subsamples have statistically consistent halo masses despite being split by X-ray temperature. This is a critical requirement for an assembly bias analysis, as any observed difference in clustering amplitude must not simply arise from differences in halo mass.

We additionally measure the cross-component of the weak lensing signal, $\Delta\Sigma_\times$, as a null test for systematic effects. The measured cross signal is consistent with zero over the full radial range, indicating the absence of significant additive shear systematics in the measurements.

Unlike some previous weak lensing analyses, we do not subtract the signal measured around random cluster positions. The primary objective of the present weak lensing analysis is to verify that the high- and low-temperature cluster subsamples possess statistically consistent halo masses. Since both subsamples are drawn from the same parent catalog and have very similar sky and redshift distributions, any additive correction associated with random points would affect both samples similarly and therefore largely cancel in the relative comparison. Likewise, we do not apply a boost-factor correction, as any dilution of the lensing signal arising from residual contamination by cluster member galaxies is expected to be very similar for the two subsamples owing to their comparable selection and redshift distributions. Consequently, neither omission affects our primary conclusion regarding the consistency of the halo masses between the two cluster subsamples.

We model the measured lensing signal using an NFW density profile \citep{NFW_profile}, fitting simultaneously for the halo mass $M_{200{\rm m}}$ and concentration parameter $c_{200{\rm m}}$. The fits are performed over the radial range $0.1 \leq R \leq 1.0\,{\rm Mpc}$ in order to minimize the impact of the two-halo term and large-scale structure contributions. The shaded regions in Fig.~\ref{fig:WL_result} represent the $1\sigma$ credible intervals of the inferred models.

For the high-temperature subsample, we obtain a best-fit halo mass of $\log\left(M_{200{\rm m}}/h^{-1}M_\odot\right)=14.33^{+0.07}_{-0.07}$ with a corresponding concentration parameter $c_{200{\rm m}} = 5.72^{+0.99}_{-0.84}$. For the low-temperature subsample, we infer $\log\left(M_{200{\rm m}}/h^{-1}M_\odot\right)=14.33^{+0.10}_{-0.09}$, and $c_{200{\rm m}} = 4.55^{+1.13}_{-0.93}$. The posterior distributions for both subsamples are shown in Fig.~\ref{fig:WL_result}. The inferred halo masses are highly consistent, with a difference $\Delta \log M_{200{\rm m}}=-0.01 \pm 0.12$, corresponding to an insignificant mass offset of approximately $0.1\sigma$. This demonstrates that the temperature split does not introduce any measurable mass bias within the current statistical uncertainties.

The quality of the fits is acceptable for both subsamples. The high-temperature sample yields a chi-squared value of $\chi^2 = 9.60$ for ${\rm dof}=4$, while the low-temperature sample gives $\chi^2 = 1.46$ for ${\rm dof}=4$. During the analysis, we see that one radial bin in the high-temperature subsample contributed disproportionately to the total chi-squared, likely due to a statistical fluctuation in the measured weak lensing signal. To investigate this, we repeated the analysis excluding this radial bin from the fit. In the main analysis shown in Fig.~\ref{fig:WL_result}, we therefore use one fewer radial bin for the high-temperature subsample compared to the low-temperature sample. Importantly, we find that the inferred halo mass constraints remain statistically consistent irrespective of whether this radial bin is included or excluded from the fit. We present the corresponding comparison in Appendix~\ref{sec:weak_lensing_appendix}.

\subsection{Cluster assembly bias}
Having established that the high- and low-temperature cluster subsamples have statistically consistent halo masses, we now investigate their relative clustering with galaxies. At fixed halo mass, a difference in the clustering amplitude of the two subsamples may indicate a dependence of halo bias on assembly history. We quantify this difference using the cluster--galaxy cross-correlation ratio.

\begin{figure}
    \centering
    \includegraphics[width=\linewidth,keepaspectratio]{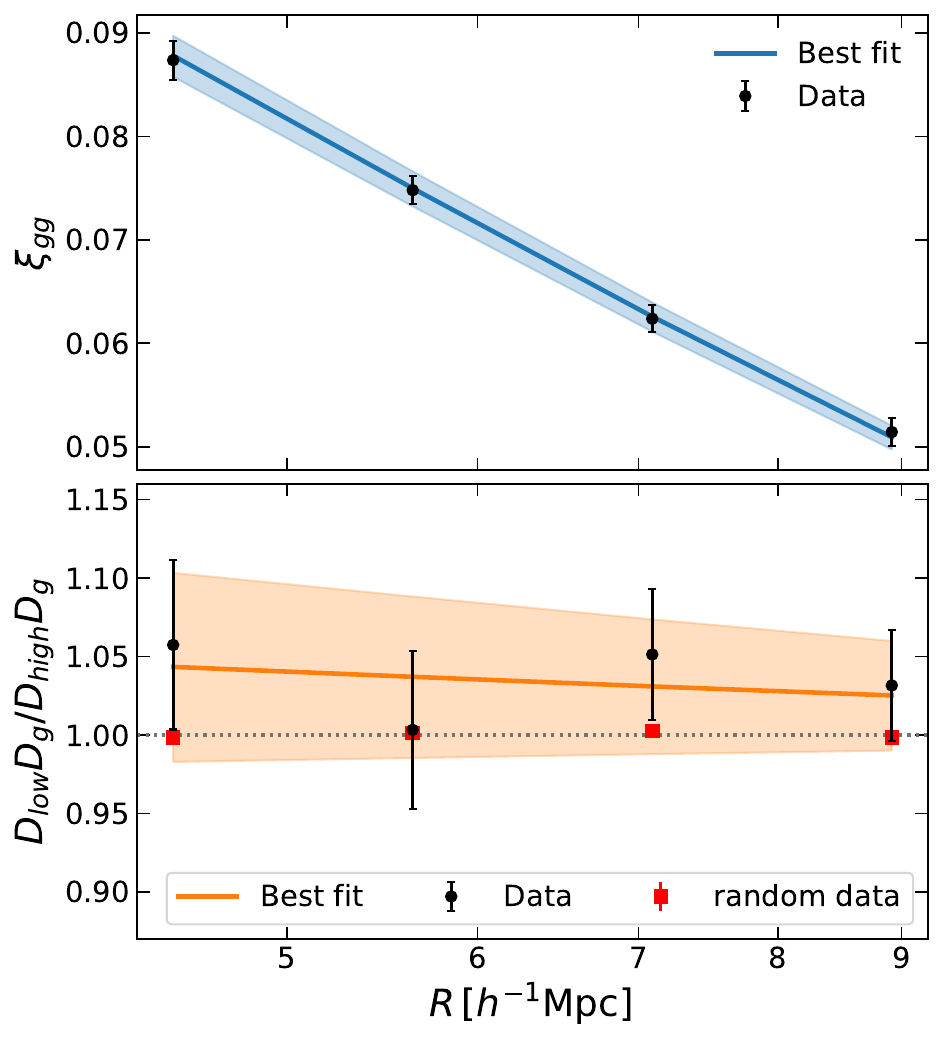}
   \caption{Best-fitting model for the joint analysis of the galaxy clustering signal and clustering ratio for the temperature-selected cluster subsamples. The top panel shows the measured clustering ratio using Eq.~\ref{eq:temp_split}, with a signal-to-noise ratio of 22.34, together with the corresponding best-fitting assembly bias model ($\Delta b$; orange solid line). The red points in the right panel show the average clustering ratio obtained from 100 independent random realizations of the temperature split, demonstrating consistency with the null expectation of unity ($p$-value=0.94). The bottom panel shows the projected galaxy correlation function, $\xi_{gg}$, with a signal-to-noise ratio of 43.11, together with the best-fitting galaxy bias model ($b_{\rm g}$; blue solid line).  In both panels, the shaded regions indicate the 68\% credible interval of the model prediction, while the black data points with error bars show the measurements used in the joint likelihood analysis. The resulting assembly bias analysis yields $\chi^2/{\rm dof}=17.10/6$ ($p$-value $=0.962$), indicating that the model provides a good fit to the data.}
    \label{fig:temp_split_result}
\end{figure}

Denoting the cluster--galaxy pair counts for the high- and low-temperature subsamples as $D_{\mathrm{high}} D_{\rm g}$ and $D_{\mathrm{low}} D_{\rm g}$, respectively, equation~\ref{eq:bias} becomes
\begin{equation}\label{eq:temp_split}
1 + \Delta b \, b_{\rm g} \, \xi_{\rm mm}
\simeq
\frac{D_{\mathrm{high}} D_{\rm g}}{D_{\mathrm{low}} D_{\rm g}} .
\end{equation}
Where $\Delta b = (b_{\mathrm{high}} - b_{\mathrm{low}})$. Figure~\ref{fig:temp_split_result} presents the joint fit to the projected galaxy clustering signal and the clustering ratio using Equation~\ref{eq:temp_split}. The uncertainties are estimated using jackknife resampling, with the data divided into 20 sky regions containing approximately equal numbers of clusters. We verify that the angular extent of each jackknife region is substantially larger than the maximum radial scale considered in our measurements. Figure~\ref{fig:temp_split_result} also shows the galaxy autocorrelation function measured using the Landy--Szalay estimator \citep{landyszalay93}, with uncertainties estimated using the same jackknife regions.

The joint fit yields a galaxy bias of ($b_{\rm g}=1.610^{+0.018}_{-0.019}$) and an assembly bias parameter of ($\Delta b=0.8\pm1.1$). The best-fitting model provides a good description of the data, with a total ($\chi^2=17.10$) for 6 degrees of freedom, corresponding to a reduced chi-square of ($\chi^2_{\rm red}=2.85$) and a $(p)$-value of 0.962. Individually, the projected galaxy clustering measurement is well described by the model, with ($\chi^2=7.80$) for 3 degrees of freedom; ($p=0.590$), while the clustering-ratio fit yields ($\chi^2=2.05$) for 2 degrees of freedom; ($p=0.960$). The corresponding signal-to-noise ratios of the projected galaxy clustering, clustering ratio, and combined analysis are 43.11, 22.34, and 46.60, respectively.

Although the best-fit value of \(\Delta b\) is positive, it is consistent with zero within the \(1\sigma\) uncertainty, indicating that the current data do not provide statistically significant evidence for cluster assembly bias. To assess the expected magnitude of the signal, we measure \(\Delta b=0.197\) in the FLAMINGO simulation for a halo sample matched in median mass and redshift to the observed sample. This value lies well within the \(1\sigma\) uncertainty of our observational measurement, indicating consistency between the measured signal and the expectation from the simulation. The details of this analysis are presented in Appendix~\ref{app:flamingo}.

As an additional consistency check, our inferred galaxy bias, $b_{\rm g}=1.610^{+0.018}_{-0.019}$, is broadly consistent with previous measurements of the large-scale bias of the DES Y3 RedMaGiC galaxy sample. In particular, \citet{Pandey2021} find $b_{\rm g}=1.74\pm0.12$, $1.82\pm0.11$, and $1.92\pm0.11$ for the three redshift bins spanning $0.15<z<0.65$. Our measurement is therefore consistent with the low-redshift RedMaGiC bias at approximately the $1\sigma$ level. A direct quantitative comparison with the individual tomographic measurements is nevertheless not strictly appropriate, since our value represents an effective bias over the full $0.15<z<0.65$ redshift range and is obtained using a different projected clustering statistic, scale range, and fiducial cosmology. In particular, the matter correlation function entering our bias model is evaluated for a slightly different cosmological model, which can lead to a modest shift in the inferred galaxy bias.

As a robustness test, we repeat the analysis after randomly assigning clusters to the two temperature subsamples while preserving the original sample sizes. This randomization is performed independently 100 times, and the resulting measurements are averaged to suppress fluctuations arising from any particular random realization. As expected in the absence of a physical assembly bias signal, the mean clustering ratio is consistent with unity at all scales ($p$-value 0.94), demonstrating that the analysis pipeline does not introduce a spurious signal through the sample splitting procedure. This provides additional confidence that any measured deviation from unity arises from the temperature-based selection rather than from the methodological biases.

\begin{figure}
    \centering
    \includegraphics[width=\linewidth]{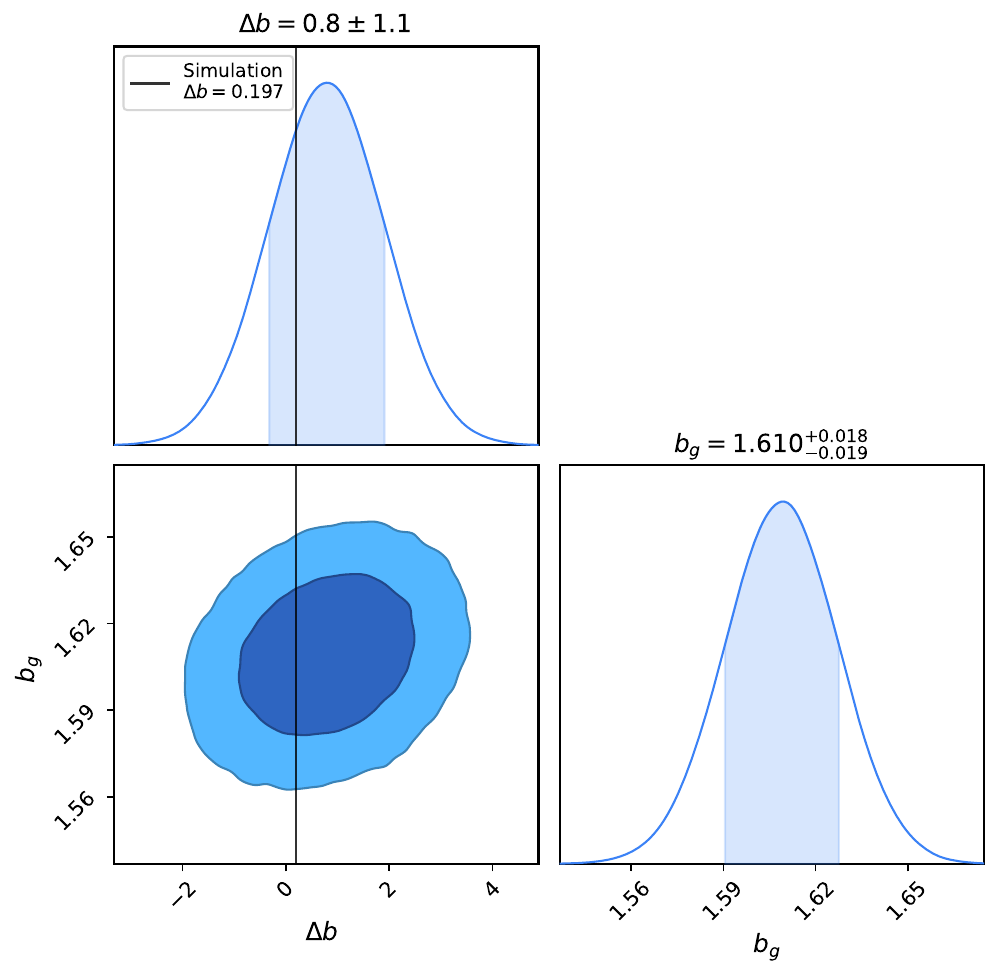}
    \caption{\textit{Posterior of bias signal}: The corner plot shows the posterior of the galaxy bias ($b_{\rm g}$) and assembly bias ($\Delta b$). The light and dark regions are the 68 and 95 credible regions. The black vertical line shows the measured expected value from FLAMINGO cosmology simulations. }
    \label{fig:bias_contour}
\end{figure}

\section{Conclusion}
\label{sec:conclusion}

The measurement of halo assembly bias requires a comparison of the large-scale clustering of halo populations while separating differences in halo mass from correlations with secondary properties. This comparison also requires careful control of observational selection effects that can differ across the survey footprint. In this work, we introduce and test a pair-count-based methodology for measuring differences in the large-scale bias of two cluster populations drawn from the same parent catalogue. Under the condition that the correlation function remains in the linear regime, the ratio of cluster--galaxy pair counts can be related directly to the difference in the linear bias of the two cluster populations, while the galaxy autocorrelation provides an independent constraint on the galaxy bias. We demonstrate the applicability of this approach using 524 X-ray selected galaxy clusters from the eROSITA eRASS1 catalogue, whose non-uniform survey exposure is explicitly accounted for when constructing the temperature-selected subsamples. We use DES Y3 RedMaGiC galaxies as the reference tracer and DES Y3 weak gravitational lensing measurements as an independent test of the halo masses of the two populations.

The main findings of our analysis are as follows:
\begin{itemize}
     
\item We verify the validity of the linear approximation underlying the pair-count-ratio methodology on the scales used in the analysis. The measured correlation function remains at $\xi(r)\ll1.0$ over $4$--$10\,h^{-1}{\rm Mpc}$, satisfying the condition $|\xi|\ll1$ required for the first-order expansion. This establishes that, on these scales, the ratio of the normalized cluster--galaxy pair counts can be used to constrain the difference in the linear bias of two cluster populations.

\item We demonstrate the methodology on a temperature-selected split of the eROSITA cluster sample. The eRASS1 survey has non-uniform exposure across its footprint, and we therefore explicitly control for this observational variation when constructing the two populations. The high- and low-temperature subsamples contain 213 and 211 clusters, respectively, and are constructed to have matched distributions in X-ray luminosity, redshift, and exposure. This matching ensures that the comparison of their large-scale clustering is performed between populations with comparable observational selection. 

\item The joint modeling of the galaxy autocorrelation and the cluster--galaxy pair-count ratio yields $b_{\rm g}=1.610^{+0.018}_{-0.019}$ and $\Delta b=0.8\pm1.1$. The corresponding $\Delta b$ is consistent with zero at the $1\sigma$ level, showing that the present data do not provide statistically significant evidence for a difference in the large-scale bias of the two temperature-selected populations.

\item As an external consistency check, our inferred bias difference is compatible with the level of assembly bias measured for analogous halo populations in the FLAMINGO simulation. Although the present data do not provide a statistically significant detection at the $1\sigma$ level, they do not rule out assembly bias at the level expected from the simulation. Our inferred galaxy bias is also consistent within the uncertainties with previous observational measurements from \citet{Pandey2021} for RedMaGiC galaxies.

\item The weak lensing measurements demonstrate that the temperature split does not introduce a measurable difference in halo mass. We infer $\log(M_{200{\rm m}}/h^{-1}M_\odot)=14.33^{+0.07}_{-0.07}$ for the high-temperature sample and $14.33^{+0.10}_{-0.09}$ for the low-temperature sample, corresponding to $\Delta\log M_{200{\rm m}}=-0.01\pm0.12$. Thus, the two populations are consistent in halo mass to the precision of the current measurements, satisfying the principal requirement for interpreting their difference in clustering as a test of assembly-dependent bias.

\item Several internal consistency tests support the robustness of the analysis. The random temperature splits produce a clustering ratio consistent with unity, indicating that the sample-splitting procedure does not itself generate a spurious clustering signal. The weak lensing cross-component is also consistent with zero, while removal of the radial bin that contributes disproportionately to the high-temperature lensing $\chi^2$ leaves the inferred halo masses statistically unchanged. Together with the matching in luminosity, redshift, and survey exposure, these tests support the stability of the comparison against the principal statistical and observational systematics considered here.

\end{itemize}

Overall, the analysis demonstrates that controlled comparisons of cluster populations can be used to measure differences in their large-scale bias without requiring the absolute cluster bias of each population to be determined independently. An important aspect of this analysis is the explicit treatment of the non-uniform exposure of the eRASS1 survey when constructing the cluster subsamples. By matching the exposure distributions, together with X-ray luminosity and redshift, we ensure that the temperature-selected populations have comparable observational selection before their clustering is compared. Applied to the eROSITA temperature-selected sample, the methodology does not yield a statistically significant assembly bias detection, with $\Delta b=0.8\pm1.1$, but provides a framework for controlled assembly-bias measurements in X-ray selected cluster samples.

In future work, the methodology can be extended to other cluster properties that may trace halo assembly history, including X-ray morphological and structural observables. Its application to the forthcoming eROSITA catalogues incorporating the second- and third-year survey data \citep{Ramos2026}, together with deeper galaxy catalogues from surveys such as Euclid \citep{euclid} and LSST \citep{lsst}, will provide substantially improved statistical precision. The larger samples will also allow the method to be tested over narrower mass and redshift ranges and for different secondary properties.

\begin{acknowledgments}
DR acknowledges
funding from the European Research Council (ERC) under the European Union’s Horizon 2020 research and innovation program (Grant agreement No. 101053992). 
This work is based on data from eROSITA, the soft X-ray instrument aboard SRG, a joint Russian-German science mission supported by the Russian Space Agency (Roskosmos), in the interests of the Russian Academy of Sciences represented by its Space Research Institute (IKI), and the Deutsches Zentrum für Luft- und Raumfahrt (DLR). The SRG spacecraft was built by Lavochkin Association (NPOL) and its subcontractors, and is operated by NPOL with support from the Max Planck Institute for Extraterrestrial Physics (MPE). The development and construction of the eROSITA X-ray instrument was led by MPE, with contributions from the Dr. Karl Remeis Observatory Bamberg \& ECAP (FAU Erlangen-Nuernberg), the University of Hamburg Observatory, the Leibniz Institute for Astrophysics Potsdam (AIP), and the Institute for Astronomy and Astrophysics of the University of Tübingen, with the support of DLR and the Max Planck Society. The Argelander Institute for Astronomy of the University of Bonn and the Ludwig Maximilians Universität Munich also participated in the science preparation for eROSITA. 

 This project used public archival data from the Dark Energy Survey (DES). Funding for the DES Projects has been provided by the U.S. Department of Energy, the U.S. National Science Foundation, the Ministry of Science and Education of Spain, the Science and Technology Facilities Council of the United Kingdom, the Higher Education Funding Council for England, the National Center for Supercomputing Applications at the University of Illinois at Urbana-Champaign, the Kavli Institute of Cosmological Physics at the University of Chicago, the Center for Cosmology and Astro-Particle Physics at the Ohio State University, the Mitchell Institute for Fundamental Physics and Astronomy at Texas A\&M University, Financiadora de Estudos e Projetos, Funda{\c c}{\~a}o Carlos Chagas Filho de Amparo {\`a} Pesquisa do Estado do Rio de Janeiro, Conselho Nacional de Desenvolvimento Cient{\'i}fico e Tecnol{\'o}gico and the Minist{\'e}rio da Ci{\^e}ncia, Tecnologia e Inova{\c c}{\~a}o, the Deutsche Forschungsgemeinschaft, and the Collaborating Institutions in the Dark Energy Survey.

The Collaborating Institutions are Argonne National Laboratory, the University of California at Santa Cruz, the University of Cambridge, Centro de Investigaciones Energ{\'e}ticas, Medioambientales y Tecnol{\'o}gicas-Madrid, the University of Chicago, University College London, the DES-Brazil Consortium, the University of Edinburgh, the Eidgen{\"o}ssische Technische Hochschule (ETH) Z{\"u}rich,  Fermi National Accelerator Laboratory, the University of Illinois at Urbana-Champaign, the Institut de Ci{\`e}ncies de l'Espai (IEEC/CSIC), the Institut de F{\'i}sica d'Altes Energies, Lawrence Berkeley National Laboratory, the Ludwig-Maximilians Universit{\"a}t M{\"u}nchen and the associated Excellence Cluster Universe, the University of Michigan, the National Optical Astronomy Observatory, the University of Nottingham, The Ohio State University, the OzDES Membership Consortium, the University of Pennsylvania, the University of Portsmouth, SLAC National Accelerator Laboratory, Stanford University, the University of Sussex, and Texas A\&M University.

Based in part on observations at Cerro Tololo Inter-American Observatory, National Optical Astronomy Observatory, which is operated by the Association of Universities for Research in Astronomy (AURA) under a cooperative agreement with the National Science Foundation.

We acknowledge the Virgo Consortium for making the FLAMINGO simulation data available. The FLAMINGO simulations were performed using the Durham Memory Intensive system managed by the Institute for Computational Cosmology on behalf of the STFC DiRAC facility.

\end{acknowledgments}

\section*{Data Availability}
The data used in this work are publicly available. The eROSITA/eRASS1 X-ray source catalogue, based on the first six months of the eROSITA all-sky survey, is described in Ref. \cite{Merloni2024}. The DES Year 3 RedMaGiC galaxy catalog and the DES Year 3 metacalibration weak-lensing shape catalog used in this analysis are publicly released as described in Refs. \cite{Pandey2021, Gatti2021}. The FLAMINGO simulation data used in this work are publicly available through the FLAMINGO data release \citep{Helly2026}.

\section{Appendixes}

\appendix

\renewcommand{\thefigure}{\Alph{section}\arabic{figure}}
\setcounter{figure}{0} 

\section{Validation of the Linear Approximation}
\label{app:linear_regime}

In our analysis, we assume that the correlation function satisfies $|\xi(r)| \ll 1$ over the radial scales of interest, which allows us to adopt a linear approximation. In this section, we explicitly validate this assumption.

Our cluster sample is drawn from the \textit{eROSITA} survey, providing an X-ray selected catalog of galaxy clusters. To measure the clustering signal, we cross-match this sample with the \texttt{redMaPPer} catalog \citep{Rykoff_etal2014}, which identifies clusters using optical galaxy overdensities. Cross-correlating these two independently selected tracers enables a more robust estimate of the large-scale clustering signal, while mitigating selection-specific systematics.

The cross-correlation function $\xi(r)$ is computed using the Landy--Szalay estimator \citep{landyszalay93}, adopting the same luminosity and redshift cuts as in the main analysis. In addition, we impose a cut on the X-ray exposure, selecting only clusters with exposure $<250$. This ensures a more uniform selection function across the survey footprint.

The exposure cut is motivated by the dependence of the effective cluster selection on the local X-ray sensitivity. Regions with higher exposure can contain lower-luminosity and consequently lower-mass clusters, which are expected to have lower large-scale bias. Restricting the sample to a more uniform exposure range therefore reduces this selection dependence. The resulting clustering amplitude is expected to provide a conservative estimate, since inclusion of additional lower-mass systems would tend to dilute rather than enhance the measured clustering signal.

\begin{figure}
    \centering
    \includegraphics[width=0.9\linewidth,keepaspectratio]{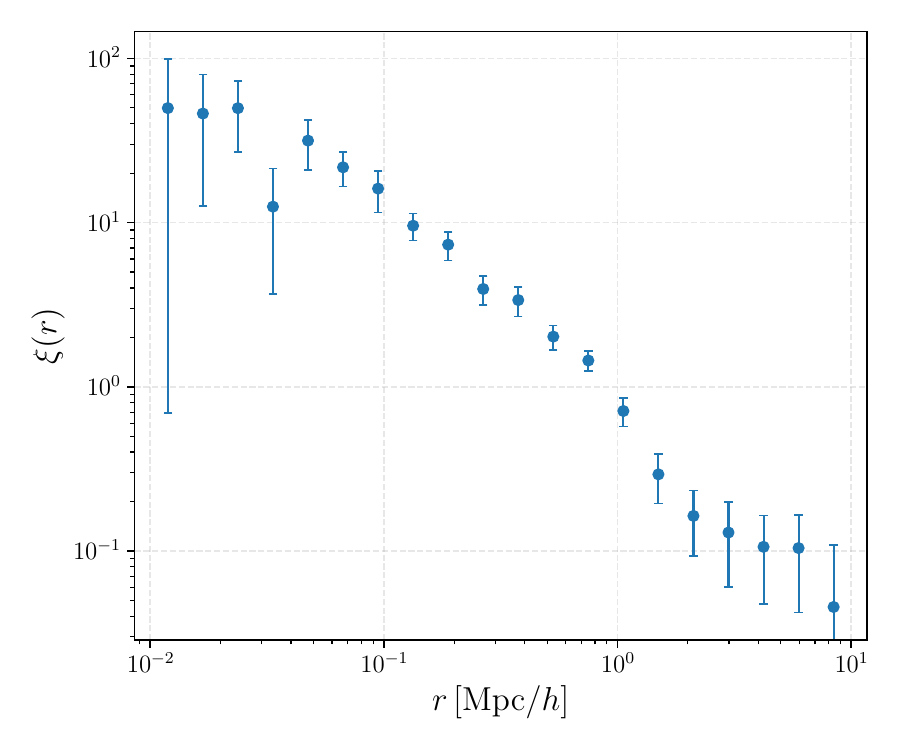}
    \caption{Cross-correlation function $\xi(r)$ between the selected cluster sample and the reference galaxy catalog, computed using the Landy--Szalay estimator. The cluster sample is selected with the same luminosity and redshift cuts as in the main analysis, with an additional exposure cut of $<250$ to ensure uniform observational conditions. We consider the radial range $4$--$10~\mathrm{Mpc}/h$, where the measured correlation amplitudes remain at $\xi(r) \lesssim 0.1$ within uncertainties. This validates the assumption $|\xi| \ll 1$ and justifies the use of the linear approximation in our analysis.}
    \label{fig:xi_validation}
\end{figure}

The resulting cross-correlation function is shown in Fig.~\ref{fig:xi_validation}. Over the radial range $4$--$10,h^{-1}\mathrm{Mpc}$, the correlation amplitude remains small, with $\xi(r)\ll1.0$ within the uncertainties. Since the first-order expansion used in the relative clustering model neglects terms of order $\mathcal{O}(\xi^2)$, the measured amplitude implies that these higher-order contributions are expected to remain small on the scales used in the analysis.

We therefore find that the condition $|\xi(r)|\ll1$ is adequately satisfied over the relevant radial range, supporting the use of the first-order approximation adopted in the relative clustering analysis.

\section{Assembly bias using FLAMINGO simulations}
\label{app:flamingo}

We use the FLAMINGO cosmological hydrodynamical simulations \citep{Schaye2023,Kugel2023} and their publicly released halo catalogues \citep{Helly2026} to assess the expected magnitude of the relative clustering difference between cluster populations selected according to their thermal properties. This provides a simulation-based reference for interpreting the relative clustering signal measured in the observed cluster sample. In particular, we use the \(L1\_m9\) simulation, which provides a large cosmological volume containing both dark-matter particles and hydrodynamical components. We use the halo-property catalogue together with the dark-matter particle distribution to construct cluster samples and measure the matter-density profiles around their centers. The halo catalogue provides the halo mass, X-ray luminosity, spectroscopic-like temperature, and halo positions used in this analysis.

To make the simulated sample representative of the observed cluster population, we first select the FLAMINGO snapshot closest to the median redshift of the observed cluster sample. We then apply the same X-ray luminosity threshold used in the observational analysis, $L_{500} > 10^{43.5}\ {\rm erg\,s^{-1}}$.

The halo population is subsequently selected to reproduce the mass scale of the observed cluster sample, with the median halo mass of the simulated sample matched to the median mass measured for the observed clusters. This ensures that the comparison is performed for halo populations with similar mass distributions, thereby reducing the contribution from the strong dependence of halo clustering on halo mass.

The selected halos are then divided into high- and low-temperature populations at fixed halo mass. Specifically, the halos are sorted into narrow mass bins, and within each mass bin the halos are divided into two approximately equal-sized populations according to their spectroscopic-like temperature. This procedure allows the clustering of halos with different thermal properties to be compared while controlling for the dominant mass dependence of halo bias. The resulting high- and low-temperature populations therefore provide a direct test of whether halos with otherwise similar masses occupy different large-scale environments.


We quantify the difference in clustering between the two simulated halo populations through their halo--matter cross-correlation functions. On sufficiently large scales, the halo--matter correlation function can be expressed in terms of the linear halo bias as
$$
\xi_{\rm hm}(r) = b\,\xi_{\rm mm}(r),
$$
where \(b\) is the linear halo bias and \(\xi_{\rm mm}(r)\) is the matter--matter correlation function. The difference in bias between the two populations can consequently be estimated from
$$
\Delta b(r)
=
\frac{
\xi_{\rm hm}^{\rm low}(r)
-
\xi_{\rm hm}^{\rm high}(r)
}{
\xi_{\rm mm}(r)
}.
\label{eq:delta_b_flamingo}
$$
For each halo population, we measure the mean dark-matter density in spherical shells centerd on the selected halos. The density contrast is defined as
$$
\delta(r) =
\frac{\rho(r)}{\overline{\rho}}-1,
$$
where \(\rho(r)\) is the mean dark-matter density in a radial shell and \(\overline{\rho}\) is the mean dark-matter density of the simulation volume. For a stack of halo positions, this density contrast provides an estimate of the halo--matter cross-correlation,
$$
\delta(r) \simeq \xi_{\rm hm}(r).
$$
We therefore calculate the bias difference directly from the difference between the stacked density-contrast profiles of the two halo populations and the matter correlation function calculated for the FLAMINGO cosmology at the redshift of the selected snapshot.

We measure the profiles in four logarithmically spaced radial bins over \(4<r<10\,h^{-1}\,{\rm Mpc}\). The resulting bias differences are
$$
\Delta b(r) =
[0.263,\;0.181,\;0.317,\;0.032],
$$
For the four radial bins, we obtain \(\Delta b(r)=[0.263,\,0.181,\,0.317,\,0.032]\), respectively. Taking the arithmetic mean of these measurements, we find a characteristic bias difference of
$$
\left\langle\Delta b\right\rangle = 0.197
$$
between the two simulated halo populations. This provides a simulation-based estimate of the level of relative clustering that can arise from differences in halo thermal properties at fixed halo mass.

\setcounter{figure}{0} 
\begin{figure}
\centering
\includegraphics[width=0.4\textwidth]{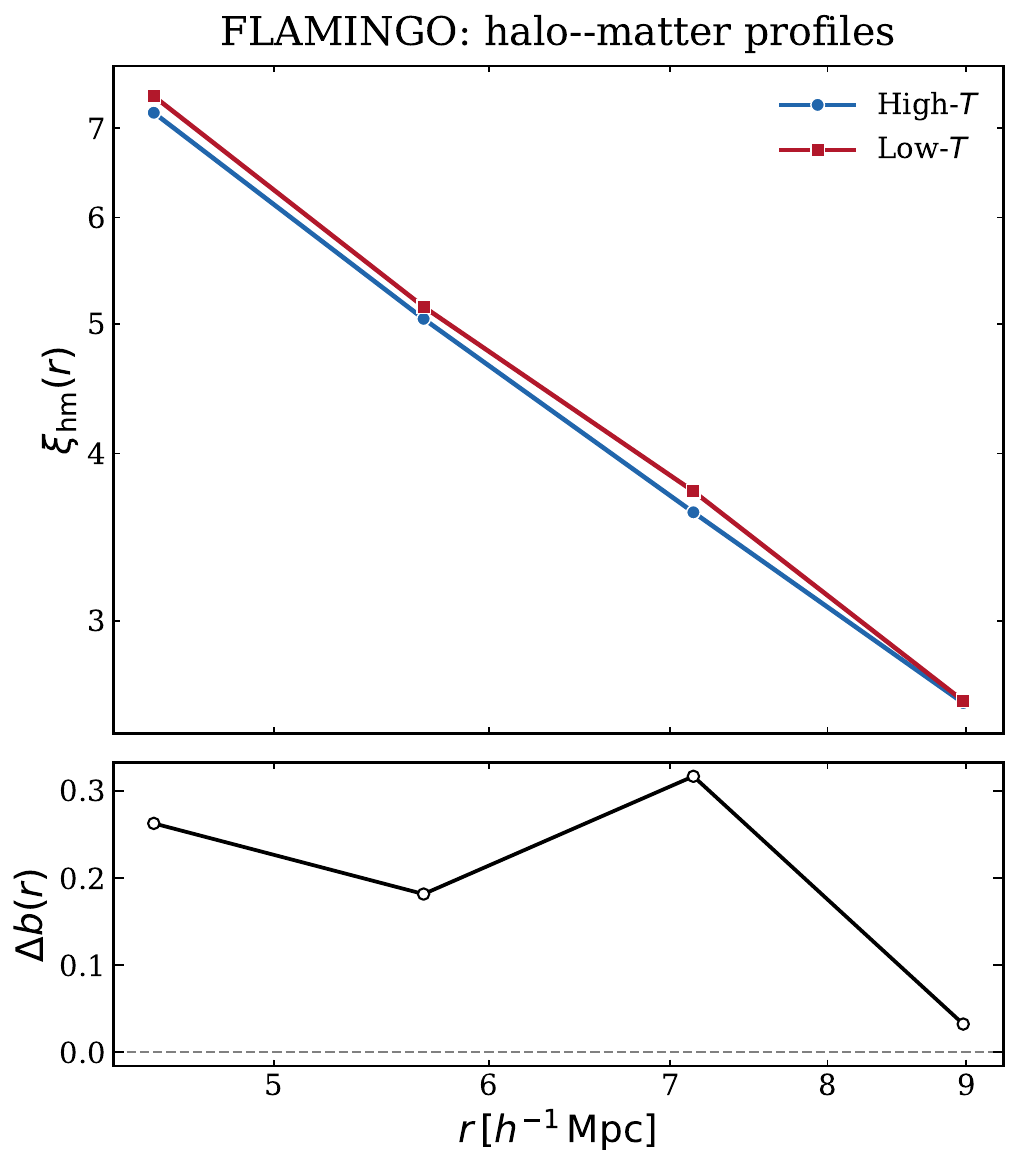}
\caption{bias difference measured in the FLAMINGO simulation. The points show \(\Delta b(r)\) calculated from the difference between the stacked halo--matter density profiles of the low- and high-temperature halo populations, divided by the matter--matter correlation function at the corresponding redshift. The four points correspond to the radial bins between \(4\) and \(10\,h^{-1}\,{\rm Mpc}\).}
\label{fig:flamingo_delta_b}
\end{figure}

\section{Mathematical model}
\label{app:mathematical_model}

In this section, we present a detailed mathematical framework for the interpretation of our clustering measurements. We begin by defining the two-point cross-correlation function, $\xi_{xy}(r)$, which quantifies the excess probability, relative to a random distribution, of finding a tracer of type $x$ at a given separation vector $\vec{r}$ from a tracer of type $y$. 

Using the continuous density contrast field $\delta(\vec{r}) = (\rho(\vec{r}) - \bar{\rho})/\bar{\rho}$, the cross-correlation function over a cosmic volume $V$ is expressed as the volume average of the product of the two density contrast fields:
\begin{equation}
\xi_{xy}(r) = \frac{1}{V} \int \delta_x(\vec{r} + \vec{r}') \delta_y(\vec{r}') d^3r'.
\end{equation}
By expanding the density contrast in terms of the local density $\rho$ and the mean cosmic density $\bar{\rho}$, this can be rewritten as:
\begin{equation}
\xi_{xy}(r) = \frac{1}{V} \int \left( \frac{\rho_x(\vec{r} + \vec{r}')}{\bar{\rho}_x} - 1 \right) \left( \frac{\rho_y(\vec{r}')}{\bar{\rho}_y} - 1 \right) d^3r'.
\end{equation}
Expanding this integral and recognizing that the volume average of $\rho/\bar{\rho}$ is unity, we simplify the expression to:
\begin{equation}
\xi_{xy}(r) = \frac{1}{V} \int \frac{\rho_x \rho_y}{\bar{\rho}_x \bar{\rho}_y} d^3r' - 1,
\end{equation}
which is commonly rearranged to isolate the integral term:
\begin{equation}
1 + \xi_{xy}(r) = \frac{1}{V} \int \frac{\rho_x \rho_y}{\bar{\rho}_x \bar{\rho}_y} d^3r'.
\end{equation}

In observational data, we typically treat one of the tracer populations (e.g., population $y$, which may represent galaxy clusters or central galaxies) as a set of $N_c$ discrete, localized sources. We can mathematically represent this localized density field as a sum of Dirac delta functions over the positions $\vec{r}_i$ of the discrete objects:
\begin{equation}
\rho_y(\vec{r}') = \sum_{i=1}^{N_c} \delta_D(\vec{r}' - \vec{r}_i).
\end{equation}
The mean density of these discrete objects in the volume is simply the total number of objects divided by the volume, yielding:
\begin{equation}
V \bar{\rho}_y = N_c.
\end{equation}

Substituting this discrete representation back into the cross-correlation integral, the volume integral over the Dirac delta functions collapses into a discrete sum over the locations of objects $y$:
\begin{equation}
1 + \xi_{xy}(r) = \frac{1}{V}  \frac{\sum_{i=1}^{N_c} \rho_x(\vec{r}_i + \vec{r})}{\bar{\rho}_x \bar{\rho}_y}.
\end{equation}
Using the relationship $V \bar{\rho}_y = N_c$, we can simplify the denominator:
\begin{equation}
1 + \xi_{xy}(r) = \frac{1}{\bar{\rho}_x} \frac{\sum_{i=1}^{N_c} \rho_x(\vec{r}_i + \vec{r})}{N_c}.
\end{equation}
The term $\frac{1}{N_c} \sum_{i=1}^{N_c} \rho_x(\vec{r}_i + \vec{r})$ represents the average density of tracer $x$ at a separation $r$ from the objects in sample $y$, which we denote as $\rho_x^{avg}$. Therefore, we obtain:
\begin{equation}
\bar{\rho}_x (1 + \xi_{xy}(r)) = \rho_x^{avg},
\end{equation}
or equivalently, tracking the total density around all $N_c$ objects:
\begin{equation}
N_c \bar{\rho}_x (1 + \xi_{xy}(r)) = \sum_{i=1}^{N_c} \rho_x(\vec{r}_i + \vec{r}).
\end{equation}

To mitigate the effects of redshift-space distortions, which perturb the apparent line-of-sight distances, it is standard practice to integrate this 3D clustering signal along the line-of-sight coordinate $\pi$, up to a maximum distance $\pi_{max}$. Integrating both sides of the equation over $\pi$ yields:
\begin{equation}
\int_{-\pi_{max}}^{\pi_{max}} d\pi N_c \bar{\rho}_x (1 + \xi_{xy}(r)) = \int_{-\pi_{max}}^{\pi_{max}} d\pi \sum_{i=1}^{N_c} \rho_x(\vec{r}_i + \vec{r}).
\end{equation}
Let us define $\Sigma_x^{avg}$ as the total projected surface density of tracer $x$. The integration of the constant $1$ gives a factor of $2\pi_{max}$, leading to:
\begin{equation}
N_c \bar{\rho}_x \left( 2\pi_{max} + \int_{-\pi_{max}}^{\pi_{max}} \xi_{xy}(r) d\pi \right) = \Sigma_x^{avg}.
\end{equation}

\vspace{0.5cm}
Next, we transition from theoretical densities to observable data pair counts. Let us consider a small solid angle $d\Omega$ around the objects in population $y$. Multiplying our projected density by $d\Omega$ gives the observable data-data pair counts, $D_x D_y$:
\begin{align}
N_c \bar{\rho}_x d\Omega \left( 2\pi_{max} + \int_{-\pi_{max}}^{\pi_{max}} \xi_{xy}(r) d\pi \right) &= \Sigma_x^{avg} d\Omega \\
&= D_x D_y.
\end{align}

We now introduce the linear deterministic bias model, which relates the cross-correlation of tracers $x$ and $y$ to the underlying matter autocorrelation function $\xi_{mm}(r)$ via their respective linear bias parameters, $b_x$ and $b_y$:
\begin{equation}
\xi_{xy}(r) = b_x b_y \xi_{mm}(r).
\end{equation}
Substituting this bias relation into our pair count equation yields:
\begin{equation}
N_c \bar{\rho}_x d\Omega \left( 2\pi_{max} + b_x b_y \int_{-\pi_{max}}^{\pi_{max}} \xi_{mm}(r) d\pi \right) = D_x D_y.
\label{eqn:pairs}
\end{equation}

A highly effective way to isolate bias difference and cancel out systematic observational effects is to take the ratio of the cross-pair counts for two distinct $y$ populations (denoted $y1$ and $y2$) correlated with the same $x$ population. Using Equation \ref{eqn:pairs}, the ratio of the pair counts is:
\begin{equation}
\frac{D_x D_{y1}}{D_x D_{y2}} = \frac{N_{c1} \bar{\rho}_x d\Omega \left( 2\pi_{max} + b_x b_{y1} \int_{-\pi_{max}}^{\pi_{max}} \xi_{mm}(r) d\pi \right)}{N_{c2} \bar{\rho}_x d\Omega \left( 2\pi_{max} + b_x b_{y2} \int_{-\pi_{max}}^{\pi_{max}} \xi_{mm}(r) d\pi \right)}.
\end{equation}
By dividing out the number of centers ($N_{c1}$ and $N_{c2}$) to normalize the pairs, and pulling out the factor of $2\pi_{max}$, we can write the normalized ratio as:
\begin{align}
\frac{D_x D_{y1} / N_{c1}}{D_x D_{y2} / N_{c2}} &= \frac{2\pi_{max} + b_x b_{y1} \int_{-\pi_{max}}^{\pi_{max}} \xi_{mm}(r) d\pi}{2\pi_{max} + b_x b_{y2} \int_{-\pi_{max}}^{\pi_{max}} \xi_{mm}(r) d\pi} \nonumber \\
&= \frac{1 + \frac{b_x b_{y1}}{2\pi_{max}} \int_{-\pi_{max}}^{\pi_{max}} \xi_{mm}(r) d\pi}{1 + \frac{b_x b_{y2}}{2\pi_{max}} \int_{-\pi_{max}}^{\pi_{max}} \xi_{mm}(r) d\pi}.
\end{align}
Assuming that the clustering terms are small, we can apply a first-order Taylor expansion $(1+\epsilon_1)/(1+\epsilon_2) \approx 1 + \epsilon_1 - \epsilon_2$. This yields:
\begin{align}
\frac{D_x D_{y1} / N_{c1}}{D_x D_{y2} / N_{c2}} &\approx 1 + \frac{b_x}{2\pi_{max}} (b_{y1} - b_{y2}) \int_{-\pi_{max}}^{\pi_{max}} \xi_{mm}(r) d\pi \nonumber \\
&\approx 1 + \frac{b_x}{2\pi_{max}} \Delta b \int_{-\pi_{max}}^{\pi_{max}} \xi_{mm}(r) d\pi.
\end{align}
Here, $\Delta b = b_{y1} - b_{y2}$ represents the difference in the large-scale biases of the two populations. Defining the normalized pair counts as $\overline{D_x D_{y}} = D_x D_y / N_c$, and leveraging the symmetry of the matter correlation function to integrate from $0$ to $\pi_{max}$, the expression further simplifies to:
\begin{equation}
 \frac{\overline{D_x D_{y1}}}{\overline{D_x D_{y2}}} \approx 1 + \frac{b_x \Delta b}{\pi_{max}} \int_{0}^{\pi_{max}} \xi_{mm}(r) d\pi.    
\end{equation} 

Finally, we connect this framework back to the standard projected correlation estimator for galaxies. The general correlation function $\xi_{xy}(R)$ at a projected distance $R$ can be estimated using the classic natural estimator, relying on the expected number of random pairs $R_x R_y$:
\begin{equation}
\xi_{xy}(R) = \frac{D_x D_y}{R_x R_y} - 1.
\end{equation}
Because the random pairs correspond to an unclustered field where the integral of $\xi$ is zero, we can define $R_x R_y$ from our previous derivations as $R_x R_y = N_c \bar{\rho}_x d\Omega (2\pi_{max})$. Following Equation \ref{eqn:pairs} and isolating the clustering term, the projected cross-correlation evaluates theoretically to:
\begin{align}
\xi_{xy}(R) &= \frac{b_x b_y}{2\pi_{max}} \int_{-\pi_{max}}^{\pi_{max}} \xi_{mm}(r) d\pi \nonumber \\
&= \frac{b_x b_y}{\pi_{max}} \int_{0}^{\pi_{max}} \xi_{mm}(r) d\pi.
\end{align}

\section{Weak lensing}
\label{sec:weak_lensing_appendix}
To assess the robustness of the weak lensing mass inference, we repeated the NFW fit after excluding the high-temperature radial bin that contributed disproportionately to the total $\chi^2$, likely due to a statistical fluctuation in the measured lensing signal. The resulting weak lensing profiles, best-fit models, and posterior distributions are shown in Fig.~\ref{fig:WL_result_1point}. Removing this single bin substantially improves the goodness of fit for the high-temperature subsample ($\chi^2=0.63$ for ${\rm dof}=3$, compared to $\chi^2=9.60$ for ${\rm dof}=4$ in the fiducial analysis), while leaving the inferred halo masses and concentrations statistically unchanged. The recovered halo mass difference remains consistent with zero, $\Delta\log M_{200{\rm m}}=-0.06\pm0.13$, indicating that the inferred mass consistency is not driven by this individual data point.
\setcounter{figure}{0} 
\begin{figure*}
    \centering
    \includegraphics[width=0.9\linewidth,keepaspectratio]{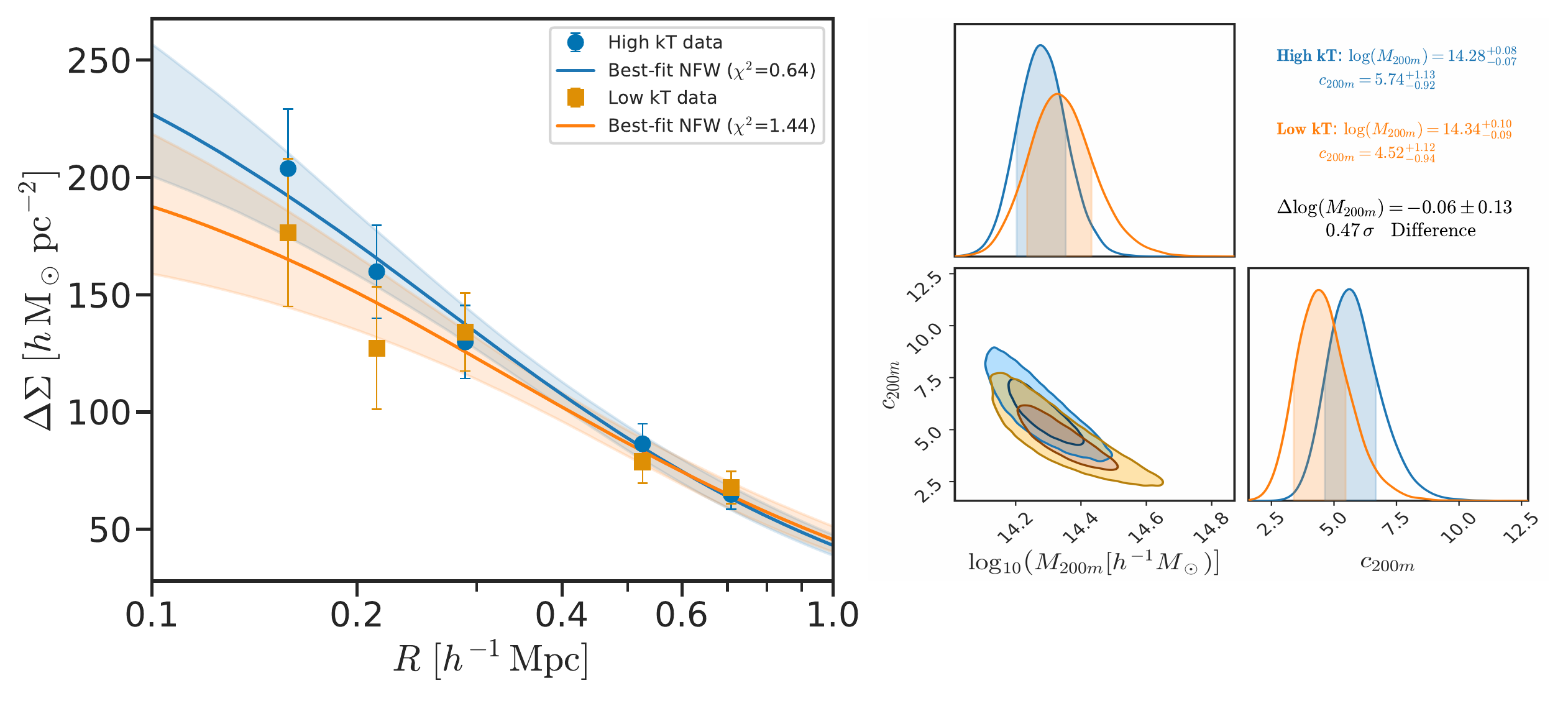}
    \caption{\textit{Weak lensing signal and posterior}: Same as Fig.~\ref{fig:WL_result}, but excluding one radial bin point that contributed disproportionately to the total $\chi^2$. The improved fit yields consistent halo mass and concentration constraints, demonstrating that the main results are robust to the removal of this single data point. }
    \label{fig:WL_result_1point}
\end{figure*}

\bibliographystyle{apsrev4-2}
\bibliography{bibliography}

\end{document}